\documentstyle[12pt]{article}
\begin{document}
\begin{center}

{\large\bf Landau Problem in Noncommutative  Quantum Mechanics}
\vskip 1cm Sayipjamal Dulat$^{a,c,}$\footnote{sdulat@xju.edu.cn} and
Kang Li$^{b,c,}$\footnote{kangli@hztc.edu.cn}\\\vskip 1cm

{\it\small$^a$ School of Physics Science and Technology, Xinjiang
University, Urumqi,
830046,China\\
$^b $Department of Physics, Hangzhou Teachers College,Hangzhou,
310036, China\\
$^c $ The Abdus Salam International Centre for Theoretical
Physics, Trieste, Italy}
 \vskip 0.5cm
\end{center}

\begin{abstract}
The Landau problem in non-commutative quantum mechanics (NCQM) is
studied. First by solving the Schr$\ddot{o}$dinger equations on
noncommutative(NC) space we obtain the Landau energy levels and
the energy correction that is caused by space-space
noncommutativity. Then we discuss the noncommutative phase space
case, namely, space-space and momentum-momentum non-commutative
case, and  we get the explicit expression of the Hamiltonian  as
well as the corresponding eigenfunctions and eigenvalues.

PACS number(s): 11.10.Nx, 03.65.-w
\end{abstract}

\section{Introduction}
Recently, there has been much interest in the study of physics on
noncommutative(NC) space\cite{SW}-\cite{Scho}, not only because
the NC space is necessary when one studies the low energy
effective theory of D-brane with B field background, but also
because in the very tiny string scale or at very high energy
situation, the effects of non commutativity of both space-space
and momentum-momentum may appear. There are many papers devoted to
the study of various aspects of quantum mechanics on
noncommutative space with usual (commutative) time coordinate.

In the noncommutative (NC) space the coordinate and momentum
operators satisfy the following commutation relations
\begin{equation}\label{eq1}
~[\hat{x}_{i},\hat{x}_{j}]=i\Theta_{ij},~~~
[\hat{p}_{i},\hat{p}_{j}]=0,~~~[\hat{x}_{i},\hat{p}_{j}]=i
\hbar\delta_{ij},
\end{equation}
where $\hat{x}_i$ and $\hat{p}_i$ are the coordinate  and momentum
operators on a NC space. Ref.\cite{zhang,Likang} showed that
$\hat{p}_i=p_i$, and  $\hat{x}_i$  have the representation form
\begin{equation}\label{eq2}
 \hat{x}_{i}=  x_{i}-\frac{1}{2\hbar}\Theta_{ij}p_{j}, \hspace{1cm} i,j =
 1,2,...,n.
\end{equation}

The case of both space-space and momentum-momentum noncommuting
\cite{zhang, Likang} is different from the case of only
space-space noncommuting.  Thus in the noncommutative (NC) phase
space the momentum operator in Eq. (\ref{eq1}) satisfies the
following commutation relations
\begin{equation}\label{eq3}
[\hat{p}_{i},\hat{p}_{j}]=i\bar{\Theta}_{ij},\hspace{2cm} i,j =
1,2,...,n.
\end{equation}
Here $\{\Theta_{ij}\}$ and $\{\bar{\Theta}_{ij}\}$  are totally
antisymmetric matrices which represent the noncommutative property
of the coordinate and momentum on noncommutative space and phase
space, respectively, and play analogous role to $\hbar$ in the
usual quantum mechanics. On NC phase space the representations of
$\hat{x}$ and $\hat{p}$ in terms of $x$ and $p$ are given in
Ref.\cite{Likang} as follows
\begin{equation}\label{eq4}
 \begin{array}{ll}
 \hat{x}_{i}&= \alpha x_{i}-\frac{1}{2\hbar\alpha}\Theta_{ij}p_{j},\\
 ~&~\\
 \hat{p}_{i}&=\alpha p_{i}+\frac{1}{2\hbar\alpha}\bar{\Theta}_{ij}x_{j}, \hspace{1cm} i,j =
 1,2,...,n.
\end{array}
\end{equation}
The  $\alpha$  here is a scaling constant related to the
noncommutativity of phase space. When $\bar{\Theta}=0$, it leads
$\alpha =1$\cite{Likang}, the NC phase space returns to the NC
space, which is extensively studied in the literature, where the
space-space is non-commuting, while momentum-momentum is
commuting.

Given the NC space or NC phase space, one should study its
physical consequences. It appears that the most natural places to
search the noncommutativity effects are simple quantum mechanics
(QM) system. So far many interesting topics in NCQM such as
hydrogen atom spectrum in an external magnetic field
\cite{nair,CST2}, Aharonov-Bohm(AB) effect \cite{CST1} in the
presence  of the magnetic field, the Aharonov-Casher effects
\cite{MZ}, and Landau problem \cite{landau}, as well as the Van de
Waals interactions and photoelectric effect in noncommutative
quantum mechanics \cite{chamoun} have been studied extensively.
The purpose of this paper is to  study  the Landau problems on NC
space and NC phase space, respectively, where
 both space-space and momentum-momentum noncommutativity could
give additional contribution.

This paper is organized as follows:  In Section 2, we study the
Landau problem on NC space. By solving the  Schr$\ddot{o}$dinger
equation in the presence of magnetic field we obtain all the
energy levels.
 In Section 3, we
investigate the Landau problem on NC phase space. By solving the
Schr$\ddot{o}$dinger equation  in the presence of magnetic field,
the additional terms related to the momentum-momentum
noncommutativity is obtained explicitly.  Conclusions are given in
Section 4.

\section{ The Landau problem on NC space}

In this section we consider the two dimensional Landau problem in
the symmetric gauge on noncommutative space. Let us consider a
charged particle, with electric charge $q$ and mass $\mu$, moving
in two dimensions (say $x-y$  plane), and under uniform magnetic
field $B$ perpendicular to the plane (say $z$ direction). The
magnetic vector potential has the form,
\begin{equation}
A_x=-\frac{1}{2}By,~~~A_y=\frac{1}{2}Bx,~~~A_z=0
\end{equation}
and then the Hamiltonian of the system has the following form,
\begin{eqnarray}\label{H-classical}
H&=&\frac{1}{2\mu}[\big(p_x^2+\frac{qB}{2c}y\big)^2+\big(p_y^2-\frac{qB}{2c}x\big)^2+p_z^2]\nonumber\\
~&=&\frac{1}{2\mu}(p_x^2+p_y^2)+\frac{1}{2}\mu\omega_L^2
(x^2+y^2)-\omega_L l_z +\frac{1}{2\mu}p_z^2,
\end{eqnarray}
where $\omega_L=\frac{qB}{2\mu c}$,  $l_z$ is the $z$ component of
the orbital angular momentum and defined as $l_z = xp_y - yp_x$.
The static Schr$\ddot{o}$dinger equation on NC space is usually
written as
\begin{equation}\label{eq6}
 H(x,p)\ast\psi = E\psi,
\end{equation}
where the  Moyal-Weyl (or star) product between two functions is
defined by,
\begin{equation}\label{eq7}
(f  \ast g)(x) = e^{  \frac{i}{2}
 \Theta_{ij} \partial_{x_i} \partial_{x_j}
 }f(x_i)g(x_j)  = f(x)g(x)
 + \frac{i}{2}\Theta_{ij} \partial_i f \partial_j
 g\big|_{x_i=x_j},
\end{equation}
here $f(x)$ and $g(x)$ are two arbitrary functions.
 On NC space
the star product can be replaced by a Bopp's shift \cite{CFZ},
i.e. the star product can be changed into the ordinary product by
replacing $H(x,p)$ with  $H(\hat{x},p)$. Thus the
Schr$\ddot{o}$dinger equation (\ref{eq6}) can be written as,
\begin{equation}\label{eq8}
H(\hat{x}_i,p_i)\psi=H(
x_{i}-\frac{1}{2\hbar}\Theta_{ij}p_{j},p_{i})\psi = E\psi.
\end{equation}
In our case, the $H(\hat x,p)$ has the following form
\begin{eqnarray}\label{eq9}
H(\hat x,p)&=&\frac{1}{2\mu}[\big(p_x^2+\frac{qB}{2c}\hat{y}\big)^2+\big(p_y^2-\frac{qB}{2c}\hat{x}\big)^2+p_z^2]\nonumber\\
~&=&\frac{1}{2\mu}\{[(1+\frac{qB}{4\hbar
c}\theta)p_x+\frac{qB}{2c}y]^2+[(1+\frac{qB}{4\hbar
c}\theta)p_y-\frac{qB}{2c}x]^2 +p_z^2\}\nonumber\\
~&=&\frac{1}{2\mu '}(p_x^2+p_y^2)+\frac{1}{2}\tilde{\mu}
\tilde{\omega}_L^2
(x^2+y^2)-\tilde{\omega}_L l_z +\frac{1}{2\mu}p_z^2\nonumber\\
~&=& H_{xy}+ H_{l_z}
 + H_\parallel\;,
\end{eqnarray}
 where
 $$
 H_{xy} =\frac{1}{2\tilde{\mu}}(p_x^2+p_y^2)+\frac{1}{2}\tilde{\mu}
\tilde{\omega}_L^2 (x^2+y^2),
\hspace{0,5cm}H_{l_z}=-\tilde{\omega}_L l_z,
\hspace{0,5cm}H_\parallel=\frac{1}{2\mu}p_z^2\;,
 $$
\begin{equation}
 \tilde{\mu}=\frac{\mu} {(1+\frac{q B}{4\hbar c}\theta)^2}\;\;, \hspace{2cm}
 \tilde{\omega}_L=\frac{q B}{2\tilde{\mu } c (1+\frac{q B}{4\hbar c}\theta)},
\end{equation}
 $H_{xy}$ is the hamiltonian for  two dimensional harmonic
oscillator with mass $\tilde{\mu}$ and angular frequency
$\tilde{\omega}_L$ . We now look for a basis of eigenvectors
common to $H_{xy}$ (eigenvalues $E_{xy}$), $H_{l_z}$(eigenvalues
$E_{l_z}$), and $H_{\parallel}$ (eigenvalues $E_{\parallel}$). It
is easy to show that the $H_{xy}$, $H_{l_z}$, and $H_{\parallel}$
commute with each other. Therefore the eigenvectors of $\{H_{xy},
H_{l_z}, H_{\parallel}\}$ will automatically be eigenvectors of
$H$ with eigenvalues
\begin{equation}
E = E_{xy} + E_{l_z} + E_{\parallel}.
\end{equation}
 The eigenvectors $\psi_k(z) \sim  e^{ik z}$ of the momentum operator
$p_z$ are also eigenvectors of $H_{\parallel}$. Thus the
eigenvalues  of $H_{\parallel}$ are of the form
\begin{equation}
E_\parallel = \frac{\hbar^2 k^2}{2\mu}, \hspace{1cm} -\infty <
k<+\infty.
\end{equation}
We see that the  spectrum of $H_{\parallel}$ is continuous, the
energy $E_\parallel$ can take  any
 positive value or zero. This result implies that $H_{\parallel}$
 describes the kinetic energy of a free particle moving along the
 $oz$( along the direction of magnetic field). The eigenfunctions
 $\psi_m (\varphi)\sim e^{im\varphi}$, $m = 0,\pm 1, \pm 2, ...$
 of $l_z$ are also wave functions of $H_{l_z}$. Therefore the
 eigenvalues of $H_{l_z}$ are
 \begin{equation}
E_{l_z} = - m \hbar \tilde \omega_L\;.
 \end{equation}
Thus, now we shall concentrate on solving the eigenvalue equation
 of $H_{xy}$ of  two-dimensional harmonic oscillator; note that the wave functions which we consider now
 depend on $x$ and $y$, and not on $z$.
 The solution to Eq.
(\ref{eq8}) can be written as a product of the solution for a
static harmonic oscillator with the phase factors responsible for
the momentum and orbital angular momentum,
\begin{eqnarray}\label{eq10}
\psi_{n_\rho m k}(\rho,\varphi,z) = R(\rho) e^{i m\varphi}e^{i k
z},\hspace{1cm} m = 0,\pm 1, \pm 2, ...,\hspace{1cm} -\infty <
k<+\infty.
\end{eqnarray}
Inserting  Eq.(\ref{eq10}) into Eq. (\ref{eq8}), and using
cylindrical coordinate system,  we can obtain the following radial
equation for the two dimensional homogenous harmonic oscillator
\begin{equation}\label{eq11}
\Big[-\frac{\hbar^2}{2\tilde{\mu}}\Big(\frac{\partial^2}{\partial\rho^2}
+ \frac{1}{\rho}\frac{\partial}{\partial\rho} -
\frac{m^2}{\rho^2}\Big) + \frac{1}{2}\mu\tilde{\omega}^2_L
\rho^2\Big]R(\rho) = E_{xy} R(\rho)\;.
\end{equation}
Solving Eq.(\ref{eq11}),  the eigenvalues of the Hamiltonian
$H_{xy}$ are
\begin{equation}\label{eq12}
E_{xy}=  (N + 1)\hbar \tilde\omega_L,
\end{equation}
with $ N=(2n_\rho +|m|), n_\rho=0,1,2,\cdots $;  and the
corresponding eigenfunctions are
\begin{equation}\label{eq13}
R(\rho) = \rho^{|m|} F(-n_\rho, |m| + 1,\zeta^2\rho^2)
e^{-{\zeta^2\rho^2}/2},\hspace{1cm} \zeta^2
=\frac{\tilde{\mu}\tilde\omega_L}{\hbar}.
\end{equation}
 Therefore the energy eigenfunctions are
\begin{equation}\label{eq14}
\psi_{n_\rho m k}(\rho\varphi,z) =\rho^{|m|} F(-n_\rho, |m| +
1,\zeta^2\rho^2)e^{i m\varphi + i k z}\;.
\end{equation}
 The
eigenvalues of the total Hamiltonian H are of the form
\begin{equation}\label{eq12}
E = (N+1)\hbar\tilde{\omega}_L - m\hbar \tilde{\omega}_L
 + \frac{\hbar^2 k^2}{2\mu}.
\end{equation}
The corresponding levels are called Landau levels. Obviously, when
$\theta =0$,  then $\tilde{\mu} \rightarrow \mu ,
\tilde{\omega}_L\rightarrow\omega_L$, our results return to the
space-space commuting case.

\section{The Landau problem on NC phase space }

The Bose-Einstein statistics in NCQM requires both space-space and
momentum-momentum non-commutativity. Thus we should also consider
the momentum-momentum non-commutativity when we deal with
 physical problems. The star product in Eq. (\ref{eq7}) on NC phase
space now is defined as
\begin{eqnarray}
(f  \ast g)(x,p) &=& e^{ \frac{i}{2\alpha^2}
 \Theta_{ij} \partial_i^x \partial_j^x+\frac{i}{2\alpha^2}\bar{\Theta}_{ij} \partial_i^p
 \partial_j^p}
 f(x,p)g(x,p)  \nonumber\\ &=& f(x,p)g(x,p)
 + \frac{i}{2\alpha^2}\Theta_{ij} \partial_i^x f \partial_j^x g\big|_{x_i=x_j}
 + \frac{i}{2\alpha^2}\bar{\Theta}_{ij} \partial_i^p f \partial_j^p
 g\big|_{p_i=p_j},
\end{eqnarray}
which can be replaced by a generalized Bopp's shift
$x_i\rightarrow \hat{x}_i, p_i\rightarrow\hat{p}_i$ with
$\hat{x}_i$ and $\hat{p}_i$ defined in Eq.(\ref{eq4}). Thus on
noncommutative phase space the Schr$\ddot{o}$dinger Eq.(\ref{eq8})
can be written as,
\begin{equation}\label{eq18}
H(\hat{x}_i,\hat{p}_i)\psi=H( \alpha
x_{i}-\frac{1}{2\hbar\alpha}\Theta_{ij}p_{j},\alpha
p_{i}+\frac{1}{2\hbar\alpha}\bar{\Theta}_{ij}x_{j})\psi = E\psi.
\end{equation}
 In two dimensions we have,
\begin{eqnarray}\label{eq19}
\hat{x}=\alpha x -\frac{\theta}{2\hbar\alpha}p_y,~~~\hat{y}=\alpha
y
+\frac{\theta}{2\hbar\alpha}p_x,\nonumber\\
\hat{p}_x=\alpha p_x
+\frac{\bar{\theta}}{2\hbar\alpha}y,~~~\hat{p}_y=\alpha p_y
-\frac{\bar{\theta}}{2\hbar\alpha}x,
\end{eqnarray}
The three parameters $\theta, \bar{\theta}$ and $\alpha $
represent the non-commutativity of the phase space, it is related
by
\begin{equation}\label{constrain}
\bar{\theta}=4\hbar^2\alpha^2 (1-\alpha^2)/\theta,
\end{equation}
so only two of them are free in the theory and they may depend on
the space and energy scales. The Hamiltonian for the two
dimensional Landau problem on noncommutative phase space in the
symmetric gauge is
\begin{eqnarray}\label{eq20}
H(\hat x,\hat{p})&=&\frac{1}{2\mu}[\big(\hat{p}_x^2+\frac{qB}{2c}\hat{y}\big)^2+\big(\hat{p}_y^2-\frac{qB}{2c}\hat{x}\big)^2+\hat{p}_z^2]\nonumber\\
~&=&\frac{1}{2\mu}\{[(\alpha+\frac{qB}{4\hbar\alpha
c}\theta)p_x+(\frac{qB}{2c}\alpha+\frac{\bar{\theta}}{2\hbar\alpha})y]^2+
[(\alpha+\frac{qB}{4\hbar\alpha
c}\theta)p_y-(\frac{qB}{2c}\alpha+\frac{\bar{\theta}}{2\hbar\alpha})x]^2 +p_z^2\}\nonumber\\
~&=&\frac{1}{2\tilde{\mu}'}(p_x^2+p_y^2)+\frac{1}{2}\tilde{\mu}'\tilde{\omega'}_L^2
(x^2+y^2)-\tilde{\omega'}_L l_z +\frac{1}{2\mu}p_z^2\nonumber\\
~&=& H'_{xy}-\tilde{\omega'}_L l_z +\frac{1}{2\mu}p_z^2,
\end{eqnarray}
where
\begin{equation}
 \tilde{\mu}'=\frac{\mu} {(\alpha+\frac{qB}{4\hbar\alpha c}\theta)^2}
\;\;,\hspace{2cm}
 \tilde{\omega'}_L=\frac{\frac{qB}{c}\alpha+\frac{\bar{\theta}}{\hbar\alpha}}{2\tilde{\mu}' (\alpha+\frac{qB}{4\hbar\alpha c}\theta)},
\end{equation}
 $H'_{xy}$ is the hamiltonian for two dimensional harmonic
oscillator with mass $\tilde{\mu}'$ and angular frequency
$\tilde{\omega'}_L$. In an analogous way as in NC space, the
solution to Eq.(\ref{eq18}) can be written as a product of the
solution for a static harmonic oscillator with the phase factors
responsible for the momentum and orbital angular momentum.
\begin{eqnarray}\label{eq22}
\psi_{n_\rho m k}(\rho,\varphi,z) = R(\rho) e^{i m\varphi}e^{i k
z},\hspace{1cm} m = 0,\pm 1, \pm 2, ...,\hspace{1cm} -\infty <
k<+\infty.
\end{eqnarray}
The eigenvalues of $l_z$ and $p_z$ are $m \hbar$ and $\hbar k$,
respectively. Choosing cylindrical coordinate system, and
inserting Eq.(\ref{eq22}) into Eq.(\ref{eq18}), we can obtain the
following radial equation for the two dimensional homogenous
harmonic oscillator
\begin{equation}\label{eq23}
\Big[-\frac{\hbar^2}{2\tilde{\mu}'}\Big(\frac{\partial^2}{\partial\rho^2}
+ \frac{1}{\rho}\frac{\partial}{\partial\rho} -
\frac{m^2}{\rho^2}\Big) + \frac{1}{2}\mu\tilde{\omega'}^2_L
\rho^2\Big]R(\rho) = E'_{xy} R(\rho)\;.
\end{equation}
This eigenvalue equation of $H'_{xy}$ leads to the wave functions
\begin{equation}\label{eq12}
R(\rho) = \rho^{|m|} F(-n_\rho, |m| + 1,\zeta'^2\rho^2)
e^{-{\zeta'^2\rho^2}/2},\hspace{1cm} \zeta'^2
=\frac{\tilde{\mu}'\tilde\omega'_L}{\hbar}.
\end{equation}
with eigenvalue
\begin{equation}\label{eq13}
E'_{xy}=  (N + 1)\hbar \tilde\omega'_L,
\end{equation}
where $ N=(2n_\rho +|m|), n_\rho=0,1,2,\cdots $.
 Therefore the total eigenfunctions of the Hamiltonian $H$ are of
 the form
\begin{equation}\label{eq14}
\psi_{n_\rho m k}(\rho\varphi,z) =\rho^{|m|} F(-n_\rho, |m| +
1,\zeta'^2\rho^2)e^{i m\varphi + i k z}
\end{equation}
where the term $e^{i k z}$ describes a free particle moving along
the magnetic field, and the particle energy is continuous. In the
x-y plane, particle is confined in a harmonic potential, energy is
discontinuous. The eigenvalues of the total Hamiltonian H are of
the form
\begin{equation}\label{eq12}
E = (N+1)\hbar\tilde{\omega'}_L - m\hbar \tilde{\omega'}_L
 + \frac{\hbar^2 k^2}{2\mu}.
\end{equation}
The corresponding levels are called Landau levels on NC phase
space. Obviously, when $\theta\neq 0 $ and $\alpha=1$, it leads to
$\bar{\theta}=0$ (refer to Eq.(\ref{constrain})), such that $
\tilde{\mu}'\rightarrow \tilde{\mu} ,~\tilde\omega'_L\rightarrow
\tilde\omega_L$, which is the space-space noncommuting case. When
both $\theta =0$ and $\bar\theta =0$ then $\tilde{\mu}'\rightarrow
\mu , \tilde{\omega'}_L\rightarrow\omega_L$, our results return to
the case of usual quantum mechanics.

\section{Conclusion }

In this letter we study the Landau problem in NCQM. The
consideration of the NC space and NC phase space produces
additional terms. In order to obtain the NC space correction to
the usual Landau energy levels, in Section 2, first, we give the
Schr$\ddot{o}$dinger equation  in the presence of a uniform
magnetic field; and then by solving the equation we derive all the
energy levels. In order to obtain the NC phase space correction to
the usual Landau problems, in Section 3, we solve the
Schr$\ddot{o}$dinger equation in the presence of a uniform
magnetic field and obtain new terms which comes from the
momentum-momentum noncommutativity.

\section{Acknowledgments} This work is supported  by the
National Natural Science Foundation of China (10465004, 10665001
and 10575026). The authors are also grateful to the support from
the Abdus Salam ICTP, Trieste, Italy.

\end{document}